\documentclass{article}

\usepackage{arxiv}

\usepackage[utf8]{inputenc} 
\usepackage[T1]{fontenc}    
\usepackage{hyperref}       
\usepackage{url}            
\usepackage{booktabs}       
\usepackage{amsfonts}       
\usepackage{nicefrac}       
\usepackage{microtype}      
\usepackage{lipsum}		
\usepackage{graphicx}
\usepackage{natbib}
\usepackage{doi}

\title{Mailing address aliasing as a method to protect consumer privacy}

\author{ \href{https://orcid.org/0000-0003-3868-8309}{\includegraphics[scale=0.06]{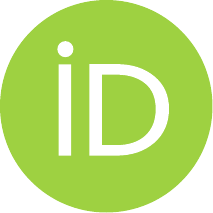}\hspace{1mm}Greg Hather} \\
    Biometrics Department\\
    Avidity Biosciences, Inc\\
    San Diego, CA, United States\\
	\texttt{ghather@gmail.com} \\
	\And
	\href{https://orcid.org/0000-0001-6594-422X}{\includegraphics[scale=0.06]{orcid.pdf}\hspace{1mm}Daniel Aranki} \\
    School of Information\\
    UC Berekeley\\
    Berkeley, CA, United States\\
	\texttt{daranki@berkeley.edu} \\
}

\renewcommand{\shorttitle}{Mailing Address Aliasing}

\hypersetup{
pdftitle={Mailing address aliasing as a method to protect consumer privacy},
pdfsubject={cs.CY},
pdfauthor={Greg Hather, Daniel Aranki},
pdfkeywords={mailing address, logistics, e-commerce, alias, privacy},
}

\begin{document}
\maketitle

\begin{abstract}
During online commerce, a customer will typically share his or her mailing address with a merchant to allow product delivery.  This creates privacy risks for the customer, where the information may be misused, sold, or leaked by multiple merchants.  While physical and virtual PO boxes can reduce the privacy risk, these solutions have associated costs that prevent greater adoption.  Here, we introduce the concept of mailing address aliasing, which may offer lower cost and greater control in some cases.  With this approach, an alias address is created that maps to the customer’s true address.  The mapping is kept private from the merchant but shared with the carrier.  We discuss the advantages and disadvantages of this approach compared with traditional methods for mailing address privacy. We find that mailing address aliasing is likely to reduce unsolicited mail to a greater extent than physical or virtual PO boxes. However, mailing address aliasing may not be compatible with all merchants' ordering systems. 
\end{abstract}


\section*{Introduction}\label{S:one}

Mail-based commerce, which has existed since the mid 1800’s (\cite{schwarzkopf2019social}), has seen enormous growth with the development of online retailing.  In 2025, global e-commerce sales are expected to exceed \$7 trillion (\cite{shettar2023commerce}).  While this growth has benefited consumers, one downside is that the customer generally discloses their residential address to the merchant to allow for product delivery.  This presents a privacy risk, where the customer’s address may be misused, sold, or leaked (\cite{pinchot2018data}).  To compound this risk, many customers share their address with multiple online merchants.  The most common consequence is unsolicited mail advertisements based on the customer’s shopping history.  In rare cases, it may lead to package theft (\cite{fung2024, fung2025}), harassment, or domestic abuse.  

Several methods for preserving privacy with package mailing are already in use.  These include PO boxes and virtual mail boxes.  However, both PO boxes and virtual mail boxes come with additional cost to the customer. In addition, PO boxes lack the convenience of home delivery.  

There are parallels between this topic and anonymity methods for electronic communications, such as email and IP addresses (\cite{edman2009anonymity}).  For electronic communication, email forwarding and anonymous email addresses can protect privacy.  For anonymity of IP addresses, Virtual Private Networks (VPNs) and Tor (\cite{dingledine2004tor}) are commonly used. Parallels also exist with online payment systems that allow customers to hide sensitive financial information from merchants (\cite{trautman2015commerce, privacydotcom}). As a notable example, the payment system PayPal serves as a trusted intermediary between the customer, the merchant, and financial institutions. However, physical package delivery has unique challenges, since using intermediate addresses can add a non-negligible cost to shipping a package. 

We stipulate that with the right technical tools, it is not required for the merchant to know the customer's shipping address for the customer to receive the goods. Rather, it is only necessary for the carrier to know the shipping address. Keeping the customer's address private from the merchant aligns with the principle of data minimization, as described in the GDPR, where data collection should be ``limited to what is necessary in relation to the purposes for which they are processed'' (\cite{GDPR2016a}). Various researchers have developed methods for data minimization, although these methods are not specific for shipping addresses (\cite{goldsteen2022data}, \cite{senarath2019data}).

\cite{bradford2001lifetime} proposed the concept of a ``lifetime address'', which is a code that maps to the receiver's true address.  The mapping is shared with the carrier but not the sender.  This approach offers some privacy advantages, but it does not reduce unsolicited mail.  In addition, the lifetime address, as proposed, does not match the format of a traditional address, and it is unlikely to work without changes to the merchant’s infrastructure.  

Here, we introduce a method to keep a customer’s mailing address private from the merchant while adding minimal cost to the online retail process. With this approach, an alias address maps to the customer’s true address, and the mapping is shared with the carrier but not the merchant. The alias address differs from the address code described in \cite{bradford2001lifetime} because the alias address matches the format of a traditional address. Additionally, permission to use alias address can expire or be revoked by the customer. Our approach, which we call mailing address aliasing, can work with no changes to most merchant’s infrastructure.  However, the approach would require some changes to the carrier’s infrastructure.  Although the method may be feasible in many countries, we focus on implementation in the United States.   

\section{Methods}

\subsection{Current process}

Prior to describing the new system, a brief overview of the current system is in order. In the United States, package delivery to consumers is dominated by 4 major carriers. Table 
\ref{tab:deliveries} shows the 2024 package volume by carrier (\cite{parcelvolpb}). 

\begin{table}
  \caption{Package volume in 2024}
  \label{tab:deliveries}
  \begin{tabular}{l c}
    \toprule
    Carrier & Packages (billions) \\
    \midrule
    USPS & 6.9 \\
    UPS & 4.7 \\
    Amazon Logistics & 6.3 \\
    FedEx & 3.7 \\
    Other & 0.8 \\
  \bottomrule
\end{tabular}
\end{table}

There are several common models for e-commerce logistics. One model is in-house logistics, where the merchant manages a warehouse, packages the product, and prints the shipping label. The merchant then gives the parcel to a carrier for final delivery (\cite{witkowski2020logistics}). Another common model is third party logistics (3PL), where a third party manages the warehouse and prints the shipping label \cite{darko2022creating}. This third party does not own the product, but instead receives a bulk delivery of the product arranged by the merchant. Note that the third party may be a major carrier that is also responsible for last-mile delivery. Alternatively, the third party could be a smaller company that brings the parcels to a major carrier for last-mile delivery. Another common model is dropshipping, where the merchant does not own or store the product. Instead, the supplier owns the product and is responsible for product storage, packaging, and label printing. Some merchants may use a mix of these logistics models. All of these models currently require the customer to share their shipping address with the merchant.  

Merchants often validate customer addresses with an API.  UPS, USPS, and FedEx all offer their own APIs to validate (\cite{upsapi, uspsapi, fedexapi}). An address validation service compares the shipping address entered by the customer against official postal databases. Generally, this database is the USPS Address Database or a database derived from it. In a 2023 survey, 53\% of online merchants use address validation of some kind (\cite{scottvalid}). Address validation can be further divided into hard validation, where the customer cannot checkout without a valid address, and soft validation, where the customer is alerted when the address cannot be validated, but the customer can still proceed with the order.   

\subsection{Proposed process}

The proposed process involves sharing the customer’s mailing address with the mail carrier but not the merchant.  For this process to work, there are several prerequisites.  First, the merchant must be a separate organization from the mail carrier, so that the address will not be shared.  Second, the merchant must not require the supplied address to be a real physical mailing address (although this prerequisite may be relaxed, as described later).  Third, the merchant must tell the customer which carrier will be used, so the customer can check if the carrier offers aliasing.  Finally, the carrier must indeed offer mailing address aliasing.  

We propose the following process for the in-house logistics model. Other logistics models, such as dropshipping or 3PL can use a modified process. 

\begin{enumerate}
\item The customer electronically sends his or her mailing address to the carrier.  In return, the carrier electronically sends the customer an alias address.  
\item The customer provides the alias address to the merchant as the shipping address.  The customer then completes the checkout process.  The merchant provides the package to the carrier with the alias address on the label.  
\item The carrier prints a new label with the mailing address and sends the package to the customer. Tracking information provided to the merchant would display the alias address rather than the mailing address.  
\end{enumerate}

\begin{figure}[ht!]
\centering
\includegraphics[width=120mm]{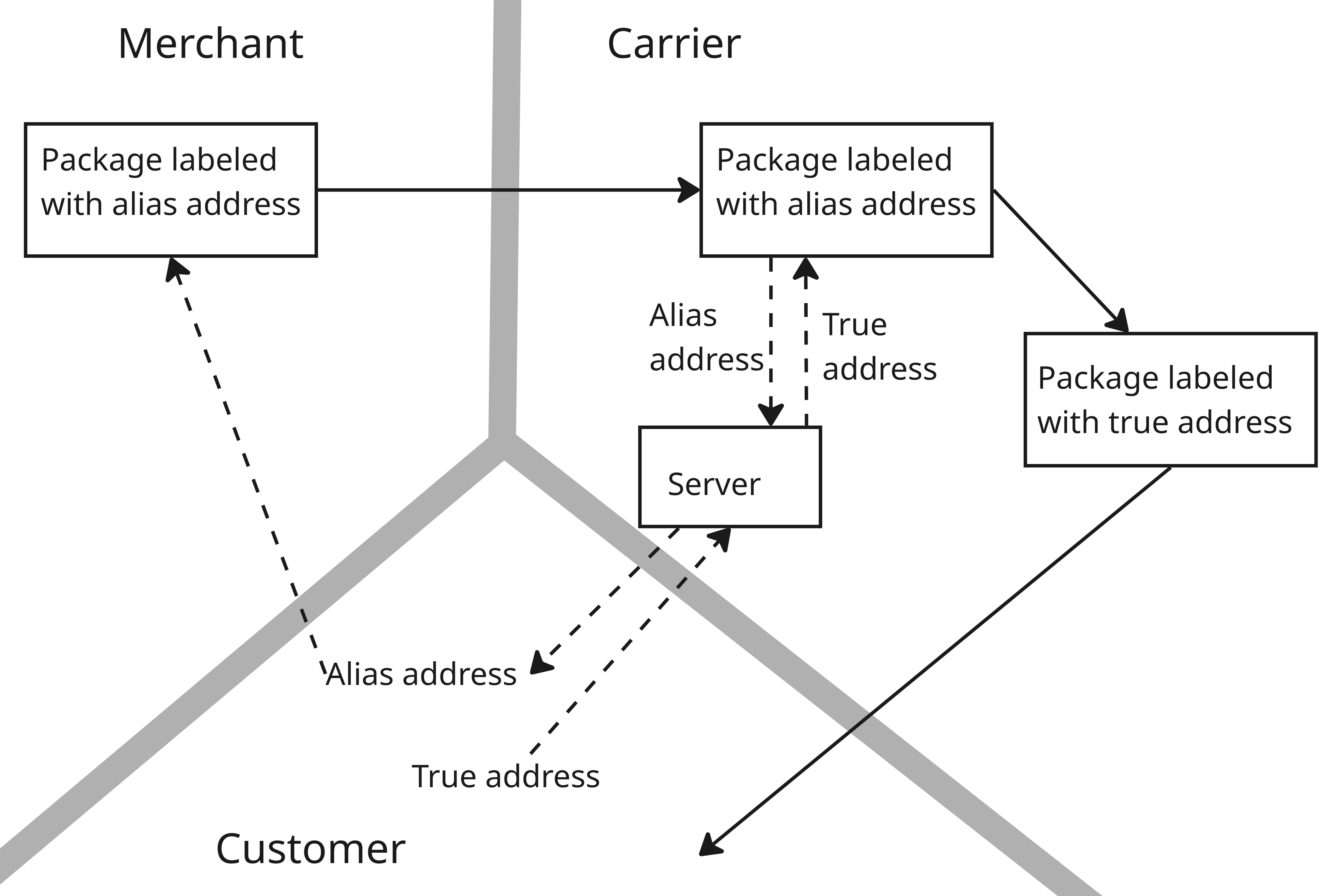}
\caption{Flow of information and physical objects. \label{flow}}
\end{figure}

Figure \ref{flow} shows the proposed process. Here, dashed lines represent the flow of information, while solid lines represent the flow of physical objects. The thick gray lines show the separation between the merchant, the carrier, and the customer. If dropshipping or 3PL companies are used, these companies can be the ones to send the package to the carrier. If the 3PL company is the same as the last mile carrier, then the process could be simplified so that merchant electronically sends the order and alias address to the carrier. In this case, the carrier would be responsible for packaging and labeling, so the first label printed would be the true address. 

Step 1 could use a website, app, or browser extension provided by the carrier.  The step could potentially occur prior to the online retail process, with the carrier sending the customer multiple alias addresses at once for later use. Step 1 could be done without any action on the part of the merchant. However, if the merchant wishes to facilitate the process, the merchant could add to their website a secure pop-up browser window connecting to the carrier's server.   

An alias address could take many forms.  Here, we consider alias address formats that would match the format of addresses in the United States.  The authors recommend that the city, state, and ZIP code remain the same between the mailing address and alias address.  Keeping the state the same is recommended because some items cannot be shipped to certain states. Keeping the city and ZIP code the same helps the merchant get accurate estimates of pricing and delivery time.  Although there is a risk of disclosure for ZIP codes with small populations, only 0.03\% of people in the United States live in ZIP codes with fewer than 100 people (\cite{census2020}).  The remaining fields to modify are then the name, address line 1, and address line 2.  All major carriers in the United States have a maximum length of no less than 30 characters for each of these fields (\cite{charlimit}).  

For example, John Smith in Any Town, New York might have a mailing address of

\begin{verbatim}
    John Smith 
    123 Main Street
    Unit 456  
    Any Town, New York, 123456
\end{verbatim}

An alias address, provided by the ABC mailing company may be

\begin{verbatim}
    ABC Alias
    2093485734 Alias Way
    Unit 8049860937
    Any Town, New York, 123456
\end{verbatim}

The alias address could be time limited, so that by agreement between the customer and the carrier, the alias address is no longer valid within a fixed number of days after first use.  The time limit can be set to reduce unwanted mail but facilitate possible returns and shipments with multiple pieces. For subscription-based orders (i.e., regularly shipped), the alias address can remain valid for the entirety of the subscription period, and canceled manually thereafter.  This could be managed by an address manager tool, akin to a password manager, which stores the web domain, the alias address, the true mailing address, and the expiration date (if any).  

Even with the limited number of characters used in the above example, with 10 digits used for the street number on Alias Way and 10 digits used for the Unit number, there are $10^{20}$ possible alias addresses for this ZIP code.  The authors suggest that a unique alias address could be used for each order.  The authors suggest that 3 of the 20 digits could be reserved to mark the year that the alias was generated, and 1 digit could be reserved for error checking.  The remaining 16 digits could be generated randomly.  To prevent accidental alias address recycling, each newly generated alias address could first be checked against a database of previously used aliases.  Since alias addresses would be distinct for each year, the carrier would only need to store the alias addresses generated in the current year (both active and inactive), as well as active alias addresses from past years.  This design could handle up to new $10^{16}$ aliases per zip code per year.  For comparison, there were 22.4 billion packages delivered in the entire US in 2024 (\cite{parcelvolpb}).  

If the merchant requires the customer to supply a physical mailing address that passes validation (i.e. hard validation), and if the API’s database does not include the alias address, then the proposed process will fail.  However, if the merchant uses the carrier’s API for address validation, the carrier can simply add the alias address to this database.  The alias address will then pass the validation step. If the merchant uses soft validation, the customer will see an alert that the address is not valid, but the customer will still be able to place the order.  

For Step 3, the carrier prints a new label with the true mailing address, which may be affixed over the original label provided by the merchant.  The new label may include a short code that is linked to the alias address in the address manager.  The customer could use this code to identify and manage the corresponding alias address within the address manager.  

Information leakage could potentially occur if the carrier is unable to deliver the package and returns it to the merchant with the customer’s true mailing address on the label.  Thus, best practice would be to remove the label from the package and replace it with a label showing the alias address. Information leakage could also occur through package tracking, since tracking information often lists checkpoints, time of delivery, and sometimes even the route taken when out for delivery.  The carrier should, at a minimum, avoid revealing the final delivery route or physical address in the tracking information.  

\section{Results}

\subsection{Implementation costs}

Printing a label is a new step for the carrier, since the label provided by the merchant would have been sufficient without aliasing.  While it may be possible to deliver the package without printing a new label, the authors recommend a new label with the true mailing address to reduce the risk of delivery errors.  In terms of cost to the carrier, adhesive thermal labels usually cost less than \$0.05. Label printers are already standard equipment at carrier stores, and the printers take only a few seconds to print a label. Thus, the per-parcel cost for mailing address aliasing is expected to be low.  

Carriers would face upfront training costs to implement mailing address aliasing. From the employee perspective, the new process is relatively simple: print a new label when alerted to do so by the carrier's package intake software. Store employees would also need to understand the system well enough to answer basic customer questions about aliasing. However, to scale the training across all store employees would require a significant investment. For example, the USPS reported having approximately 121,000 retail associates in 2020 \cite{uspsemploy}. Even a 15 minute training for all retail associates, at a rate of \$20 per hour, would cost \$605,000.

For a large carrier, the software backend to support mailing address aliasing would need to integrate with existing logistics IT systems. In addition, user interfaces would need to be created for both customers (to create the alias address) and employees (to help print a new label). Without details of the carrier's operations, obtaining an accurate development cost projection is difficult. However, we estimate that the up-front cost of software development would be in the low hundreds of thousands of dollars. 

\subsection{Comparison to related methods}

Compared to traditional methods of protecting mailing address privacy, such as PO boxes, mailing address aliasing may offer lower cost and greater convenience.  However, this approach may not work in all cases, such as when the carrier doesn’t offer address aliasing.  Thus, a customer may still keep a PO box but have packages sent to her home when aliasing is allowed.   

In addition to providing customers with more control, mailing address aliasing will give customers greater knowledge of how their personal information is being shared.  For example, if an alias address with no expiration date was used to send unwanted mail, the customer could determine which merchant shared the information.  This would discourage merchants from sharing aliased mailing addresses with other parties.  

Table \ref{tab:comparison} compares mailing address aliasing with alternatives. As shown, virtual mailboxes can forward parcels to a home address, but the customer must approve and pay for each parcel forwarded after it arrives at the virtual mailbox. Aliasing could only work for carriers that offer the service, while virtual mailboxes can receive packages from all carriers. PO boxes can only receive packages delivered by USPS. Virtual mailboxes and PO boxes have limited storage space, whereas the aliasing approach sends the package directly to the customer's home address, which usually has fewer space limitations. By providing aliases that have time limited validity, mailing address aliasing can help block unsolicited mail. In addition, since each order uses a different alias address, there is a reduced chance that the data could be coupled (that is, connected) across orders or merchants. In this sense, mailing address aliasing offers the strongest privacy.   

\begin{table}
  \caption{Comparing mailing address aliasing with a virtual mail and a PO box}
  \label{tab:comparison}
  \begin{tabular}{l c c c}
    \toprule
    Attribute & Aliasing & Virtual mailbox & PO box \\
    \midrule
    Able to send parcels to home address & Yes & Yes & No  \\
    Requires customer input to forward & No & Yes & No  \\
    Can receive packages from all carriers & No & Yes & No \\
    Space limitations & No & Yes & Yes \\
    Limits data coupling & Yes & No & No \\
    Limits unsolicited mail & Yes & No & No \\
    Additional costs involved & Yes & Yes & Yes \\
  \bottomrule
\end{tabular}
\end{table}

\section{Discussion}

\subsection{Incentives for carriers and merchants}

There are several incentives for the different parties to participate in mailing address aliasing. For the carrier, address aliasing may be a way of adding value to the service. The carrier could even make aliasing a premium service, charging the customer a small fee per parcel or per month. The carrier could even partner with privacy focused organizations to raise awareness and find early customers. 

While merchant action is not necessarily required for mailing address aliasing, having buy-in from the merchant makes the system more likely to work. For the merchant, there may be a greater chance of making a sale if the customer knows she can protect her address. Offering greater privacy may even be a brand advantage for merchants (\cite{jakobi2021data}). Thus, stronger customer privacy may boost trust and potentially sales.

\subsection{Additional considerations}

We note that aliasing would not work with Amazon, since Amazon is both a merchant and a package delivery service.  

The carrier may face reduced revenue if address aliasing reduces unsolicited mail delivered by the same carrier.  For example, USPS may have a disincentive to provide mailing address aliasing, since 19\% of its revenue in 2023 was from marketing mail (\cite{uspsrev}).  However, USPS and other carriers may offset loss of revenue and other costs by seeking payment from the customer when she requests an alias address.  
Additionally, the carrier may lose revenue if the agreement with the customer prevents the carrier from selling the customer’s personal data.  

We anticipate several main obstacles to the implementation and adoption of mailing address aliasing. One obstacle is that carriers may be reluctant to provide up-front investment to create a service for which demand is uncertain. Another obstacle is the lack of consumer awareness about mailing address aliasing, although this may be addressed as described later. An additional hurdle is the fact that aliasing may not be compatible with all merchant websites, such as those that perform hard address validation. 

Consumer education would be an important early step as tools for mailing address aliasing are developed. Carriers could introduce customers to the process on their websites or stores. Additionally, merchants who are supportive of mailing address aliasing could link to an explanation of the process on their websites. Finally, privacy-focused consumer groups and individuals could share information about the concept with consumers.   

To be adopted at scale, the address aliasing process must be easy to use.  Common standards and shared tools developed by the major carriers or open source communities may help. In addition, a user feedback mechanism could allow carriers to collect useful information about the customer experience. Such information could yield valuable insights, enabling the system to be improved over time.

\section{Conclusion}

As part of normal modern life, a person’s home address must be shared with many parties, including utility companies, hospitals, and governments.  Complete control over this information seems unrealistic at present.  However, mailing address aliasing has the potential to improve privacy in some cases by limiting the distribution of a customer’s true address during online commerce.  

Future research could investigate how interested consumers are to keep their mailing address private when ordering online. 
Additional research could also be conducted with a focus on merchants. Merchants could be surveyed to determine what fraction are currently set up to allow mailing address aliasing and what fraction would be supportive of aliasing. Further work could also be done to build a proof-of-principle alias generation tool. Finally, exploration of mailing address aliasing in contexts other than e-commerce may be fruitful. 

Overall, mailing address aliasing offers a new privacy-focused way to conduct e-commerce. The method would benefit customers by hiding the customer address from the merchant, thus reducing the chance of unsolicited mail, and more generally, reducing the amount of information that data brokers can obtain. Whether such a system would eventually be adopted by carriers, customers, and merchants remains to be seen.  

\section*{Conflict of Interest Statement}

The authors declare that the research was conducted in the absence of any commercial or financial relationships that could be construed as a potential conflict of interest.

\section*{Author Contributions}

GH: Conceptualization, Methodology, Formal analysis, Writing – original draft. DA: Conceptualization, Methodology, Writing – review \& editing

\section*{Funding}
This research received no specific grant from any funding agency in the public, commercial, or not-for-profit sectors.

\section*{Data Availability Statement}
The dataset analyzed for this study can be found in the 2020 census 118th congressional district summary file. [https://data.census.gov/].

\bibliographystyle{unsrtnat}
\bibliography{frontiers_revised}

@article{schwarzkopf2019social,
  title={The social embeddedness of marketing},
  author={Schwarzkopf, Stefan},
  journal={The Oxford Handbook of Consumption},
  pages={27},
  year={2019},
  publisher={Oxford University Press}
}

@article{shettar2023commerce,
  title={E-Commerce: An Empirical Study.},
  author={Shettar, Rajeshwari M},
  journal={Productivity},
  volume={39},
  number={2},
  year={2023}
}

@article{pinchot2018data,
  title={Data Privacy Issues In The Age of Data Brokerage: an Exploratory Literature Review.},
  author={Pinchot, Jamie and Chawdhry, Adnan A and Paullet, Karen},
  journal={Issues in Information Systems},
  volume={19},
  number={3},
  year={2018}
}

@article{fung2024,
 author  = {Fung, Esther},
 year    = {2024},
 title   = {Porch Pirates Are Stealing {AT\&T} {iPhones} Delivered by {FedEx}},
 journal = {The Wall Street Journal}
}

@article{fung2025,
 author  = {Fung, Esther},
 year    = {2025},
 title   = {How Bribes Helped a Crime Ring Steal Thousands of {iPhones} From Porches},
 journal = {The Wall Street Journal}
}

@incollection{bradford2001lifetime,
  title={Lifetime Addresses: A New Postal Paradigm for the 21st Century},
  author={Bradford, Camille and Mayer, Jack},
  booktitle={Future Directions in Postal Reform},
  pages={365--374},
  year={2001},
  publisher={Springer}
}

@article{edman2009anonymity,
  title={On anonymity in an electronic society: A survey of anonymous communication systems},
  author={Edman, Matthew and Yener, B{\"u}lent},
  journal={ACM Computing Surveys (CSUR)},
  volume={42},
  number={1},
  pages={1--35},
  year={2009},
  publisher={ACM New York, NY, USA}
}

@inproceedings{dingledine2004tor,
  title={Tor: The second-generation onion router.},
  author={Dingledine, Roger and Mathewson, Nick and Syverson, Paul F and others},
  booktitle={USENIX security symposium},
  volume={4},
  pages={303--320},
  year={2004}
}

@misc{census2020,
  author = {{US Census Bureau}},
  title = {2020 Census 118th Congressional District Summary File},
  howpublished= {\url{https://data.census.gov/}},
  year = {2020}
}

@misc{charlimit,
  author = {Elrgdawy, Fatime Jr.},
  title = {Character Limits in Address Lines for {USPS}, {UPS} and {FedEx}},
  howpublished = {\url{https://www.serviceobjects.com/blog/character-limits-in-address-lines-for-usps-ups-and-fedex/}},
  year = {2020}
}

@misc{uspsrev,
author = {USPS},
title = {{US Postal Service} Reports Fiscal Year 2023 Results},
howpublished = {\url{https://about.usps.com/newsroom/national-releases/2023/1114-usps-reports-fiscal-year-2023-results.htm}},
year = {2023}
}

@article{trautman2015commerce,
  title={E-Commerce, cyber, and electronic payment system risks: lessons from PayPal},
  author={Trautman, Lawrence J},
  journal={UC Davis Bus. LJ},
  volume={16},
  pages={261},
  year={2015},
  publisher={HeinOnline}
}

@incollection{jakobi2021data,
  title={Data Privacy: A Driver for Competitive Advantage},
  author={Jakobi, Timo and von Grafenstein, Max and Schildhauer, Thomas},
  booktitle={The Machine Age of Customer Insight},
  pages={147--158},
  year={2021},
  publisher={Emerald Publishing Limited}
}

@misc{GDPR2016a,
  year       = {2016},
  title      = {Regulation ({EU}) 2016/679 of the {European} {Parliament} and of the {Council}},
  url        = {https://data.europa.eu/eli/reg/2016/679/oj},
  author     = {{European Parliament} and {Council of the European Union}},
}

@article{goldsteen2022data,
  title={Data minimization for GDPR compliance in machine learning models},
  author={Goldsteen, Abigail and Ezov, Gilad and Shmelkin, Ron and Moffie, Micha and Farkash, Ariel},
  journal={AI and Ethics},
  volume={2},
  number={3},
  pages={477--491},
  year={2022},
  publisher={Springer}
}

@article{senarath2019data,
  title={A data minimization model for embedding privacy into software systems},
  author={Senarath, Awanthika and Arachchilage, Nalin Asanka Gamagedara},
  journal={Computers \& Security},
  volume={87},
  pages={101605},
  year={2019},
  publisher={Elsevier}
}

@misc{parcelvolpb,
  year       = {2025},
  title      = {Parcel shipping index report 2024},
  url        = {https://www.pitneybowes.com/content/dam/pitneybowes/us/en/shipping-index/dmr-2334-parcel-shipping-index-ebook-finalv3.pdf},
  author     = {{Pitney Bowes}},
}

@article{witkowski2020logistics,
  title={Logistics models in E-commerce},
  author={Witkowski, Krzysztof and Koralewska, Monika and Huk, Katarzyna},
  journal={Vedeck{\'e} Pr{\'a}ce Materi{\'a}lovotechnologickej Fakulty Slovenskej Technickej Univerzity v Bratislave so S{\'\i}dlom v Trnave},
  volume={28},
  number={46},
  pages={90--97},
  year={2020},
  publisher={De Gruyter Poland}
}

@article{darko2022creating,
  title={Creating valuable relationships with third-party logistics (3PL) providers: a multiple-case study},
  author={Darko, Eric Owusu and Vlachos, Ilias},
  journal={Logistics},
  volume={6},
  number={2},
  pages={38},
  year={2022},
  publisher={MDPI}
}

@misc{uspsapi,
  author = {USPS},
  year = {2025},
  title = {Addresses 3.0 | devportal},
  howpublished = {\url{https://developer.usps.com/addressesv3}}
}

@misc{upsapi,
  author = {UPS},
  year = {2025},
  title = {UPS Developer Portal},
  howpublished = {\url{https://developer.ups.com/tag/Address-Validation}}
}

@misc{fedexapi,
  author = {FedEx},
  year = {2025},
  title = {Address Validation API},
  howpublished = {\url{https://developer.fedex.com/api/en-us/catalog/address-validation.html}}
}

@misc{scottvalid,
  author = {Edward Scott},
  year = {2023},
  title = {Have an Address Validator},
  howpublished = {\url{https://baymard.com/blog/address-validator}}
}

@misc{uspsemploy,
  author = {USPS},
  year = {2020},
  title = {Retail details},
  howpublished = {\url{https://news.usps.com/2020/03/03/retail-details/}}
}

@misc{privacydotcom,
  author = {Privacy.com},
  year = {2025},
  title = {Virtual cards to protect your payments},
  howpublished = {\url{https://www.privacy.com/virtual-card}}
}






\end{document}